\documentstyle[preprint,aps]{revtex}
\begin{document}
\draft
\preprint{DOE/ER/40427-15-N94, TMU-NT940801}
\title{
Photon  Vector-Meson Coupling and Vector Meson Properties
\\at Low Temperature Pion Gas}
\author{
Su Houng Lee$^{1,2}$\cite{slee},
Chungsik Song$^{3}$\cite{song} and
Hiroyuki Yabu$^{4}$\cite{yabu} }
\address{
 $^1$Department of Physcis, Yonsei University, Seoul 120-749, Korea \\
 $^2$Department of Physics, FM-15,
  University of Washington, Seattle, WA 98195, USA \\
 $^3$Cyclotron Institute, Texas A\&M University,
   College Station, TX 77843, USA \\
 $^4$Department of Physics, Tokyo Metropolitan University, Hachioji, Tokyo
   192, Japan}

\date{August 10 1994}

\maketitle

\begin{abstract}
The vector and axial vector current mixing phenomena
at low temperature pion gas by Dey, Eletsky and Ioffe, leads to the
low temperature correction of the photon-vector meson coupling ($g_\rho$)
at order $\epsilon=T^2/6f_\pi^2$ and the $\rho$  meson mass
at  order $\epsilon^2$.
We show how this {\it low temperature theorems} involving the photon and vector
mesons are satisfied in the chiral models based on
hidden gauge symmetry and the  massive Yang-Mills approach with an explicit
$a_1$ meson.
We discuss possible phenomenological consequences of the low temperature
corrections in RHIC experiments.
\end{abstract}

\newpage


 For many years, dilepton spectrum from Relativistic Heavy Ion Collision (RHIC)
has attracted a lot of interest\cite{RHIC}.
This is so because once produced, they will predominantly
escape from the collision region without further interaction carrying
information about the hot initial stages of the collision.
Of particular interests are the vector meson resonance regions because
model calculations show definite relations between chiral symmetry
restoration at finite temperature or density to vector meson
masses.
If this is so, chiral symmetry restoration could be directly observable
through shift of vector meson peak positions in the dilepton spectrum.

The vector meson properties and dilepton spectrum at  finite temperature
have been studied in various ways \cite{Pisa,DEI,GKS,Boch,FHL,HKL,HL}.
Recently, it was realized by Dey et.al. [DEI] \cite{DEI} that in the
lowest order in $\epsilon=T^2/6f_\pi^2$, where $f_\pi=93$ MeV,
there is no change in the vector meson masses and
only mixing between vector and axial vector
correlator takes place.
They obtained the result based only on the PCAC and current algebra.
The low temperature limit of the QCD sum rule calculations at finite
temperature indeed satisfies this constraint\cite{HKL,Eletsky2}.

In the effective model approach,
chiral perturbation theory\cite{Wein,GL84}, although successful in reproducing
systematically the low energy hadron physics, is not able to reproduce the
higher energy region including the resonances.  Therefore,
effective chiral models with explicit $\rho$ and $a_1$ meson degrees of freedom
were used to calculate the vector meson properties and dilepton spectrum
  from the low temperature pion gas\cite{GKS}.
This way, although we can not generate
the $\rho$ meson pole, the changes of the vector meson properties due to
low temperature pionic effect could be obtained in a chirally invariant way.
In this respect, at low temperature,
the effective model approaches should also be consistent with the
mixing phenomena while the $\rho$  meson mass changes
at the order $\epsilon^2$.

The purpose of this letter is the following.  First, we will
summarize the mixing phenomena by DEI and its implications to the
effective photon vector-meson coupling and the vector meson mass.
Second,  we will show  how
the correct temperature dependence of the photon vector meson coupling
and the vector meson mass come out in the effective chiral lagrangians.
Finally, we will comment on possible physical consequences in RHIC.


Let us consider the thermal average of an operator ${\cal O}$:
\begin{eqnarray}
\langle {\cal O} \rangle_T=\sum_n \langle n| {\cal O} exp(-H/T)|n \rangle
\end{eqnarray}
Here the sum is over the full set of eigenstates of the hamiltonian $H$.
 At low temperature ($T < m_\pi$), we shall take into account only the vacuum
state and the lowest excitation mode of the hadron gas, the pions.
Then the thermal average can be approximated by
\begin{eqnarray}
\label{pion}
\langle {\cal O} \rangle_T=
\langle {\cal O} \rangle_{T=0}+
\sum_{a=1}^3 \int { d^3 p \over  2 \omega (2\pi)^3 }
\langle \pi^a(p)| {\cal O} |\pi^a(p) \rangle
n_B(\omega/T),
\end{eqnarray}
where $\omega=\sqrt{p^2+m_\pi^2} $, $a$ denotes the isospin index and
$n_B(x)=[e^x-1]^{-1}$ is the Bose-Einstein distribution.  We have used the
covariant normalization for the state vector here
$\langle \pi^a(p)|\pi^b(p') \rangle=2 \omega (2 \pi)^3 \delta^{ab} \delta^3
(p-p')$.

Eq.(\ref{pion}) contains the first non-trivial term of the expansion by the
pion number density at $T \neq 0$ and is a good approximation up to
$T \sim$ 150 MeV\cite{GL,LS}.  For
larger $T$, the effect of higher powers of the pion density and other
massive excitations $(K,\eta,...)$ start to become important.  The matrix
element can be evaluated in the soft pion limit such that
\begin{eqnarray}
\label{soft}
\langle \pi^a(p)| {\cal O} | \pi^a(p) \rangle=-\frac{1}{f_\pi^2}
\langle 0| [ {\cal F}_5^a [ {\cal F}_5^a ,{\cal O} ]]|0 \rangle
+O({m_\pi^2}),
\end{eqnarray}
where ${\cal F}_5^a$ is the axial charge operator.
Eq.(\ref{pion}) with eq.(\ref{soft}) when applied to chiral order
parameter,
gives the leading temperature
correction $\langle \bar{q} q \rangle_T=\langle \bar{q} q \rangle(1-
\frac{3}{4} \epsilon )$, with $\epsilon=T^2/6f_\pi^2$ for SU(2)  and
as given by the low temperature
theorem of chiral perturbation in ref \cite{GL}.

Let us now look at the time ordered vector correlator at finite temperature
\begin{eqnarray}
\label{eq0}
\Pi^V_{\mu \nu}(q,T) = i \int d^4 x e^{iqx} \langle T[ V_\mu^a(x) V_\nu^a(0)]
\rangle_T.
\end{eqnarray}
The vector current is $ V^a_\mu=\bar{q} \gamma_\mu \frac{\tau^a}{2} q$ with
$\tau^a$ being the SU(2) isospin matrix normalized as ${\rm Tr}[\tau^a \tau^b]
=2 \delta^{ab}$.
DEI\cite{DEI}
showed that in the leading order in $\epsilon$, $\Pi^V_{\mu\nu}$ mixes with the
corresponding axial correlator  $\Pi^A_{\mu\nu}$, such that
\begin{eqnarray}
\label{eq1}
\Pi^V_{\mu \nu}(q,T) &=& (1-\epsilon)\Pi^V_{\mu \nu}(q,0)+\epsilon
                    \Pi^A_{\mu \nu}(q,0)  \\ \nonumber
\Pi^A_{\mu \nu}(q,T) &=& (1-\epsilon)\Pi^A_{\mu \nu}(q,0)+\epsilon
                    \Pi^V_{\mu \nu}(q,0).
\end{eqnarray}
This mixing appears only in isospin triplet channel.
For the correlator of isospin singlet vector current $V^\omega_\mu=
\frac{1}{2}(\bar{u} \gamma_\mu u+\bar{d} \gamma_\mu d)$, however,
 there is no effect to this order;
\begin{eqnarray}
\label{eq2}
\Pi^{SV}_{\mu \nu}(q,T) = \Pi^{SV}_{\mu \nu}(q,0).
\end{eqnarray}
This is so because two pions can not couple to $\omega$ current.

Let us now look at the imaginary part of the vector correlators near the
$\rho,\omega$ meson resonance region.   Due to the mixing effect, the
strength of the $\rho$ peak is reduced while the pole position
is not changed.  The reason for the later is due to the vanishing of the
 $\rho-\pi$ scattering length in the soft pion limit (Adler's zero).
The change of the
residue of the $\rho$ meson pole can be effectively understood by the change of
the coupling constant ($g_\rho$) of the vector current to
$\rho$ meson  at finite temperature.
In the resonance region, as can be seen in Fig.~1.~a,
the external current coupling to the $\rho$ meson is reduced by
thermal pions
\begin{eqnarray}
\label{gr}
g^2_\rho(T)=(1-\epsilon)g^2_\rho+ O(\epsilon^2).
\end{eqnarray}
Looking at the imaginary part near the $a_1$ meson resonance,
we also note that the vector current will couple to the axial mesons
to lowest order in $\epsilon$ (Fig.~1.~b).
This interpretation is also
consistent with the origin of the mixing.   The interaction of two pions with
one current decreases the contributions of the initial channel.  Admixture of
opposite parity channel arises when the  two pions interact with two currents
at
points $0$ and $x$.

It should be emphasized that the mixing effect is a consequence of chiral
symmetry and current algebra so that it
has to be satisfied in the low temperature limit of any model calculation;
i.e. when only the lowest thermal pion contributions are taken into account
in the thermal average.


In the literature, there are two known methods of introducing the vector
mesons and the external photon field into the chiral lagrangian.  These are
the massive Yang-Mills approach (MYMA) and the hidden gauge approach (HGA).
In the MYMA,  the $\rho$ and  the $a_1$ are
introduced as external gauge fields of the chiral group
and the photon field is
introduced via VMD\cite{Sak}.  In the HGA, the vector meson is introduced
as the gauge field of the hidden local symmetry and the photon is
introduced as the external gauge field\cite{BKY}.
These two methods have been shown to be gauge
equivalent\cite{Yamawaki1,Zahed1} and give identical results\cite{Yamawaki2}
if both the $\rho$ and $a_1$ are included.

Starting from an effective lagrangian, the lowest order
temperature corrections are obtained by considering the thermal pion loops.
In this respect, it is similar to calculating the loop correction in chiral
perturbation theory,
but since we are interested in the temperature correction only,
no infinities occur and no additional higher derivative terms are needed.
The additional momentum factor comes in with the temperature and gives
correction to the tree level parameters.


Let us consider the chiral lagrangian of Bando et.al.\cite{BKY},
which is based on the
[SU(2)$_{\rm L} \times$SU(2)$_{\rm R}]_{\rm global} \times$
[SU(2)$_V]_{\rm local}$ ``linear" sigma model.  It is constructed with two
SU(2)-matrix valued variables $\xi_L(x)$ and $\xi_R(x)$, which transforms
as $\xi_{L,R}(x) \rightarrow \xi'_{L,R}(x)=h(x) \xi_{L,R}~ g^\dagger_{L,R}$
under
$h(x)\in[$SU(2)$_V]_{\rm linear}$ and $g_{L,R}\in[$SU(2)$_{L,R}]_{\rm global}$.
The vector meson $V_\mu$  is introduced as the gauge field of the local
symmetry  and
the photon ${\cal B}_\mu$  as the external gauge field of the global
symmetry.   Then the general invariant lagrangian looks as follows,
\begin{eqnarray}
{\cal L} & = & {\cal L}_A+ a{\cal L}_V+{\cal L}_{\rm kin}
         ( V_\mu ,{\cal B}_\mu),  \\ [12pt] \nonumber
{\cal L}_V & = & f_\pi^2 {\rm tr} \left[V_\mu-\frac{1}{2i} ( {\cal D}_\mu \xi_L
         \cdot
         \xi_L^\dagger + {\cal D}_\mu \xi_R \cdot \xi_R^\dagger)  \right]^2,
  \\ [12pt] \nonumber
{\cal L}_A & = & f_\pi^2 {\rm tr} \left[\frac{1}{2i} ( {\cal D}_\mu \xi_L \cdot
         \xi_L^\dagger - {\cal D}_\mu \xi_R \cdot \xi_R^\dagger) \right]^2.
\end{eqnarray}
Here ${\cal L}_A$ is precisely the gauged CCWZ lagrangian \cite{CCWZ}
and
\begin{eqnarray}
{\cal D}_\mu \xi_{L,R}= \partial_\mu \xi_{L,R}+ ie \xi_{L,R} {\cal B}_\mu
 \tau_3/2.
\end{eqnarray}
In the ``unitary" gauge,
\begin{eqnarray}
\xi_L^\dagger(x)=\xi_R(x)=e^{i\pi(x)/f_\pi} \equiv \xi(x)
\end{eqnarray}
and rescaling $V_\mu \rightarrow gV_\mu$, the effective lagrangian takes
the form,
\begin{eqnarray}
{\cal L} & = & -\frac{1}{4}(F^{(V)}_{\mu \nu})^2 -\frac{1}{4}
     (\partial_\mu {\cal B}_\nu- \partial_\nu {\cal B}_\mu)^2
     +\frac{1}{4} {\rm tr}(\partial_\mu U \partial^\mu U^\dagger)
     +\frac{1}{2} m_\rho^2 V_\mu^2-eg_\rho V_3^\mu {\cal B}_\mu
     +\frac{1}{2} m_{\cal B} {\cal B}_\mu^2  \cr
   & &+
      g_{\rho \pi \pi} V^\mu \cdot (\pi \times \partial_\mu \pi)
      +g_{\gamma \pi \pi}   {\cal B}^\mu  (\pi \times \partial_\mu \pi)_3
      +{\cal L}_{>3}
\end{eqnarray}
where the parameters are given as $m_\rho^2=ag^2 f_\pi^2$,
$g_\rho=a g f_\pi^2$, $ g_{\rho \pi \pi}=\frac{1}{2}a g $,
$g_{\gamma \pi \pi}=(1-\frac{1}{2}a)e $.  For $a=2$ ,
these formula are well known to
automatically give the universality of the $\rho$-couplings, KSRF  relations
(KSRF I : $g_\rho=2 f_\pi^2 g_{\rho \pi \pi}$, KSRF II; $m_\rho^2=2
g_{\rho \pi \pi}^2 f_\pi^2$) and the $\rho$ meson dominance of the
electromagnetic form factor of the pion ($g_{\gamma \pi \pi}=0$).

First, let us examine the consistency of the hidden local symmetry approach
with the {\it low energy theorem}.
We need to calculate  the low temperature correction to $g_\rho$.
This can be obtained by calculating the one-loop correction
and considering the thermal correction coming from pions
only\cite{GL}.
In general there are two types of loops.  The first tadpole type of
contribution can be obtained by looking at the higher non-derivative
pion terms appearing in ${\cal L}_{>3}$. Its temperature effect can be
easily calculated by forming a pion tadpole loop and inserting the temperature
dependent part of the thermal pion propagator  $iD_\pi^T(p)=iD_\pi^{T=0}(p)+
2 \pi n_B(p_0) \delta(p^2)$.  This leads to the following simple
formula for the correction to a general interaction term involving the
$\pi,\rho,{\cal B}$
\begin{eqnarray}
\label{ted}
f(\pi,V,{\cal B}) \longrightarrow [1+\epsilon (d/d\pi^a)^2]
f(\pi,V,{\cal B}).
\end{eqnarray}
This clearly breaks chiral symmetry because the pion loop expansion is not
consistent with the chiral symmetry, but this breaking is in the next
higher order and we need not mind it in the lowest-order calculation.
For the non-tadpole type of diagrams, each loop has to be calculated
separately.

There is a temperature dependent
correction for the photon-$\rho$ meson coupling by the tadpole type of
diagram considered in Fig.~1.~a. Its contribution can be easily obtained
by looking at the photon-$\rho$-$\pi$-$\pi$ coupling and using eq.~(12);
\begin{equation}
g_\rho(T)=g_\rho(1-{\epsilon\over2}).
\end{equation}
For $\omega$ meson one can explicitly see that such a coupling cancels.
The $\rho$ meson mass is modified by the diagram shown in Fig.~2.
However, its contribution is of
order $\epsilon^2$ because of the presence of the derivative
in the $\rho\pi\pi$ coupling.
We obtain the correct low temperature dependence of $g_\rho$ and $m_\rho$.
This is a crucial low temperature theorem that any model with explicit
vector meson has to satisfy.   In the extended version with explicit $a_1$
meson, there exist an additional photon-$a_1$-$\pi$ coupling, so that the
$a_1$ meson pole also appears as in eq.(\ref{eq1})


Now let us discuss how the {\it low temperature theorem} is satisfied in
the MYMA.  The Lagrangian looks as follows.

\begin{eqnarray}
\label{MYMA}
 {\cal L}={1 \over 4}f_\pi^2 {\rm tr} \left[ D_\mu U D^\mu U^\dagger  \right]
 +{1 \over 2} m_\rho^2 {\rm tr}
 \left[A_\mu^L A^{\mu L}+A_\mu^R A^{\mu R} \right]
\end{eqnarray}
where,
\begin{eqnarray}
D_\mu U=\partial U-igA_\mu^L U +igUA_\mu^R
\end{eqnarray}
and $A_\mu^L=V_\mu+A_\mu,A_\mu^R=V_\mu-A_\mu$
and $U=exp[2i\pi/f_\pi]$.  This with the vector meson dominance, the
external vector field $V_\mu$ will be identified with the vector meson fields
by $V_\mu=g_\rho \rho_\mu$.
The relevant terms needed to determine the  change in rho meson properties
are,
\begin{eqnarray}
{\cal L}& = & g^2f_\pi \epsilon^{abc}\pi^a \rho_b^\mu a_c^\mu+ {g^2 \over 2}
  \left[ \rho_\mu^2 \pi^2- \rho^\mu \cdot \pi \rho_\mu \cdot \pi \right]
  \nonumber \\ [12pt]
  & & + {1 \over 2} (f_\pi^2 g^2 +m_\rho^2)a_\mu^2 +
{1 \over 2} m_\rho^2 \rho_\mu^2 + \cdot \cdot \cdot
\end{eqnarray}
Here  the $a_1$ mass is given by $m_a^2=m_\rho^2+f_\pi^2 g^2 $
(Weinberg relation).
The main difference with the HGA is that here there is a $\rho \rho \pi
\pi$ coupling which  at first sight gives a change in the $\rho$ meson mass
to order $\epsilon$.   However, this contribution has to be added by the
$\rho$ meson self energy with an intermediate $a_1$ and $\pi$.
Together with the rho meson propagator, these contribute to
eq.(\ref{eq0}) as follows to leading order in $\epsilon$

\begin{eqnarray}
\Pi^V_{\mu \nu}(q,T) & = & ( g_{\mu \nu}-{q_\mu q_\nu \over m_\rho^2} )
g_\rho^2
\left[ {i \over m_\rho^2-q^2}  +{i \over m_\rho^2-q^2} ig^2f_\pi^2
\epsilon {i \over m_\rho^2-q^2}
 - {i \over m_\rho^2-q^2} {i^2 g^4 f_\pi^4 \epsilon  \over m_a^2-q^2}
{i \over m_\rho^2-q^2} \right]  \nonumber
 \\ [12pt]
 & = & ( g_{\mu \nu}-{q_\mu q_\nu \over m_\rho^2} )
g_\rho^2 \left[ (1-\epsilon)  {i \over m_\rho^2-q^2}   +
 \epsilon  {i \over m_a^2-q^2} \right].
\end{eqnarray}
Again vector correlator shows mixing phenomena at finite temperature
and satisfies the {\it low temperature theorem}.
However, it should be noted that in this, if we did not include the $a_1$
meson, we would have erroneously obtained a $\rho$ meson mass shift to
order $\epsilon$.   Whereas in the HGA,
the {\it low temperature theorem} for the $\rho$ meson
are satisfied with only the $\rho$ included.
This is so because in the HGA, the $\rho$ is a gauge particle
of the hidden gauge group and independent of the external
left- and right- chiral symmetry, whereas in the MYMA, $\rho$ and $a_1$ is
related via chiral transformation.


Let us discuss thermal dilepton production from a pion gas\cite{Rush}.
The electromagnetic current is defined as,
\begin{eqnarray}
J^{em}_\mu=V^\rho_\mu+\frac{1}{3} V_\mu^\omega-\frac{\sqrt{2}}{3} V_\mu^\phi.
\end{eqnarray}
The relevant quantity is the thermal average of the
retarded correlator of the electromagnetic
current, which is related to the time ordered correlator by
${\rm Im} [\Pi^F_{\mu \mu}]={\it tanh}^{-1}(E/2T) {\rm Im }[\Pi^R_{\mu \mu}]$.
In the soft pion limit, this is a function of only the invariant mass and
therefore, the
thermal dilepton emission rate per invariant mass can be written as
\begin{eqnarray}
\label{dilepton}
{ dR \over d M^2}= \alpha_{em}^2 4 T^2 {\rm Im}[\Pi^F(M^2)] B_1(M/T)
\end{eqnarray}
where $B(y)=\frac{1}{2 \pi^2} \int_y^\infty dx \sqrt{x^2-y^2}/(e^y+1)$.
Due to mixing, the imaginary part of the time ordered correlator is given by,
\begin{eqnarray}
\label{im1}
{\rm Im} [\Pi^V(M^2,T)]&=&(1-\epsilon) { f_\rho^2 \Gamma_\rho m_\rho \over
                    (M^2-m_\rho)^2 + \Gamma_\rho^2 m_\rho^2 } +
                         {\rm continuum}  \\ \nonumber
{\rm Im} [\Pi^{SV}(M^2,T)]&=& { f_\omega^2 \Gamma_\omega m_\omega \over
                    (M^2-m_\omega)^2 + \Gamma_\omega^2 m_\omega^2 } +
                         {\rm continuum}
\end{eqnarray}
in the soft pion limit near the resonance region.
Here, for simplicity, we have used a Breit-Wigner parameterization and
$m_\rho=770 $ MeV, $\Gamma_\rho=150 $ MeV, $f_\rho=152$ MeV and
$m_\omega=782 $ MeV, $\Gamma_\omega=9.8 $ MeV, $f_\omega=46$ MeV.

In principle, the ratio of peak height between the $\rho$ meson and
$\omega$ meson should decrease as $1-\epsilon$.  In Fig.~3, we have plotted
the expected thermal dilepton production eq.(\ref{dilepton}) at $T$=100
and 150 MeV, above which the leading order expression should break down.
The origin of the relative reduction of the
$\rho$ peak in the dilepton spectrum is coming from the medium dependent
decay constant, which
is related to its $e^+ e^-$ partial width as
$g_\rho^2=3 m_\rho \Gamma(\rho \rightarrow e^+ e^-)/4 \pi \alpha^2$.
Because of this reason,  even $\rho $ meson produced in the initial stage
of the collision will also be affected as long as its decay into dileptons
takes place inside the thermal pion gas.

In previous experiment, it was not possible to separate  $\rho$ and
$\omega$ peaks in the dilepton spectrum of RHIC\cite{new}.
Also, it is not clear how
the leading temperature effects should modify at higher temperature.
However, present hadronic model calculations to understand the NA 38 data
at CERN\cite{NA38}  on $J/\Psi$ suppression and $\phi/(\rho
+\omega)$ enhancement in O+U and S+U collision is consistent with an overall
reduction of $\rho +\omega$ observed in the dilepton spectrum.
Koch, Heinz and Pi\u{s}\'{u}t \cite{Heinz1}
first obtained a  relation between the transverse
energy $(E_T)$ and the fireball size or impact parameter by fitting it to the
$J/\Psi$ suppression with increasing $E_T$.  The suppression comes mainly
 from $J/\Psi$ absorption.  Then assuming that the ratio of $\phi/\omega$
production from rescattering is the same as in the primary collisions,
using the fact that $\phi$ absorption cross section is less than that of
the $\omega$ and the same relation between $E_T$ and
impact parameter, they were able to explain the $\phi$ enhancement data.
However, taking into account secondary $\rho$ production, the fit
deteriorates unless there is an additional mechanism to reduce  the
rho meson decaying into dileptons\cite{Heinz1,Heinz2}.
In fact, the original data are even consistent with a systematic
reduction of $N_\rho+N_\omega$ in A A collision compared to $N_\rho+N_\omega$
in p A collision times A\cite{NA38}.


In summary, we have shown that both effective chiral model
of hidden gauge approach and massive Yang-Mills approach are
consistent with the mixing properties of the vector current correlator
at low temperature.
We have also discussed
some phenomenological consequence expected in
the dilepton channel in the $\rho$  and $\omega$ resonance region.


We would like to thank M. Lutz for emphasizing the equivalence between the
HGA and MYMA.  We would like to thank T. Hatsuda, C.M. Ko, G. A. Miller
M. Rho for useful comments.
The work of SHL was supported in part by the Basic Science Research Institute
Program, Ministry of Education of Korea, no. BSRI-94-2425 and The US
Department of Energy.
The work of CS was supported by the National Science Foundation
under Grant No. PHY-9212209 and Welch Foundation under Grant No. A-1110.
SHL and  HY would like to thank  B. Fridman and W. Noerenberg for inviting
to the ECT* workshop on "Mesons and Baryons in Hadronic Matter"
where part of this work was done and for useful discussions.

\newpage

\centerline{{\bf Figure Captions}}

\vspace{1.5cm}

\noindent
{\bf Fig.\ 1} One-loop corrections to correlators in the thermal pion gas.
 Solid line denotes the zero temperature vector correlator and
 double solid line does the axial vector correlator.
 Dashed lines correspond to thermal pions.

\vspace{1cm}

\noindent
{\bf Fig.\ 2} One loop diagram for the $\rho$ meson self-energy.

\vspace{1cm}

\noindent
{\bf Fig.\ 3} Thermal dilepton production rate in $fm^{-4}$GeV$^{-2}$.
The solid line is the result obtained with low temperature correction.

\end{document}